\documentclass[usenatbib]{mn2e}
\usepackage{natbib}
\usepackage{hyperref}
\hypersetup{
    pdfnewwindow=true,      
    colorlinks=true,       
    linkcolor=blue,          
    citecolor=black,        
    filecolor=blue,      
    urlcolor=blue           
}
\usepackage{graphicx}
\usepackage{aas_macros}
\begin{document}

\title[GRB Precursor Catalog]{Catalog of Isolated Emission Episodes in Gamma-ray Bursts from Fermi, Swift and BATSE}

\author[M. Charisi, S. Marka, I. Bartos]{M.~Charisi,$^{1}$\thanks{mc3561@columbia.edu} S.~M\'arka,$^{2}$ I.~Bartos,$^{2}$\\
$^1$Department of Astronomy, Columbia University, New York, NY 10027, USA \\
$^2$Department of Physics, Columbia University, New York, NY 10027, USA}

\maketitle

\begin{abstract}
We report a comprehensive catalog of emission episodes within long gamma-ray bursts (GRBs) that are separated by a quiescent period during which gamma-ray emission falls below the background level. We use a fully automated identification method for an unbiased, large scale and expandable search. We examine a comprehensive sample of long GRBs from the BATSE, Swift and Fermi missions, assembling a total searched set of 2710 GRBs, the largest catalog of isolated emission episodes so far. Our search extends out to [$-1000\,\mbox{s},\,750\,\mbox{s}$] around the burst trigger, expanding the covered time interval beyond previous studies and far beyond the nominal durations ($T_{90}$) of most bursts. We compare our results to previous works by identifying pre-peak emission (or precursors), defined as isolated emission periods prior to the episode with the highest peak luminosity of the burst. We also systematically search for similarly defined periods \emph{after} the burst's peak emission. We find that the pre-peak and post-peak emission periods are statistically similar, possibly indicating a common origin. For the analyzed GRBs, we identify 24\,$\%$ to have more than one isolated emission episode, with 11\,$\%$ having at least one pre-peak event and 15\,$\%$ having at least one post-peak event. We identify GRB activity significantly beyond their $T_{90}$, which can be important for understanding the central engine activity as well as, e.g., gravitational-wave searches.
\end{abstract}


\section{Introduction}

Gamma-ray bursts (GRBs) exhibit complex temporal evolution, which is unique for each GRB. While gamma rays are produced in relativistic outflows far from the central engine that powers the jet, the observed temporal variation of the emission is likely connected to the activity of the central engine itself (e.g., \citealt{1997ApJ...490...92K}). Interpreting the temporal variation of the emission may therefore be essential in understanding the origin of GRBs and their environments.

GRB emission often does not appear to be continuous. In many cases, episodes of emission are separated by quiescent intervals during which the gamma-ray flux drops below the observed background level. 
The reason for this separation is not currently clear. It could be due to (i) different mechanisms being responsible for the separate emission periods, (ii) a temporary halt in the activity of the central engine, or (iii) a continuous central engine with greatly varying output that can, for periods, fall below the detectable level, e.g., due to modulation of the velocity of the relativistic outflow \citep{2001MNRAS.320L..25R,2001MNRAS.324.1147R}.

A number of authors have aimed to characterize the statistical properties of distinct emission episodes. Many of them focused on the so-called \emph{precursor} activity, often defined as emission preceding the episode that has the highest peak intensity (main event), from which it is separated by a quiescent period with no detectable gamma-ray flux. The search for such precursors is motivated in part by theoretical models predicting a distinct emission episode prior to the main burst. However, the models typically predict one separate episode and cannot account for the more than two emission periods found for some GRBs. (For a brief review of theoretical models for precursor emission see \citealt{2008ApJ...685L..19B}.) 

\citet{1995ApJ...452..145K} [hereafter K95] searched for precursors in a sample of GRBs detected by BATSE in the first 3 years of its operation. They required that a precursor is followed by a quiescent interval at least as long as the duration of the main burst. They found precursor activity in $\sim$3\% of the GRBs. They identified significant correlation between the duration of the precursor and the duration of the main event, although this correlation was not confirmed by the subsequent studies. Beyond this correlation, they concluded that the characteristics of the main emission episode are independent of the presence of the precursor.

Later, \citet{2005MNRAS.357..722L} [hereafter L05] relaxed the requirement for minimum quiescence and searched for precursors the flux of which decay before the onset of the main event. He looked for dim emission that did not trigger the GRB detector in a sample of long and bright GRBs detected by BATSE, and identified precursors in $\sim$20\% of the sample. He found precursor spectra to be softer than those of the main bursts.

\citet{2008ApJ...685L..19B} [hereafter B08] looked for precursors in GRBs detected by Swift that had known redshifts. They detected precursors in $\sim$15\% of them and found no significant difference between the spectral properties of precursors and main events. Subsequently, \citet{2009A&A...505..569B} [B09] searched for precursor activity in the entire BATSE catalog and studied their spectral evolution. They identified precursors in $\sim$12\%, and found that their properties are similar to those of the main emission episode. They concluded that precursor and main emission episodes probably arise from the same fireball mechanism.

More recently, \citet{2010ApJ...723.1711T} focused on a small sample of short GRBs, as well as the subset of short GRBs with extended emission, detected by Swift. They identified precursor activity in $\sim$10\% of the short GRBs, with precursor and main emission showing no substantial difference. They discussed a number of distinct emission mechanisms that could result in precursor emission preceding a short GRB.

Other authors studied isolated emission episodes to examine the activity of the central engine during quiescent periods. For instance, \cite{2001MNRAS.324.1147R} suggested that GRB afterglow observations could be used to determine whether the central engine was dormant during these periods. \cite{2001MNRAS.320L..25R} found that the duration of a quiescent interval and the duration of the subsequent emission episode are correlated, but found no correlation with the emission prior to the quiescent time. They associated this correlation with a meta-stable energy \emph{build-up} i.e. the longer this energy accumulates, the more energy is available to produce the burst following the quiescent time.

In this paper, we present a search for isolated emission episodes over a set of 2710 GRBs. We employ a fully automated search that identifies emission episodes in the time-frequency domain, allowing the analysis of a large set of bursts, while avoiding human bias.  We use detections by the Fermi, Swift and BATSE missions to maximize our data set and to allow for comparisons with previous results. We make the detailed emission-episode catalog publicly available for further analysis\footnote{\url{http://geco.markalab.org/GRBprecursors}}. We further examine and describe some of the statistical properties of the episodes.

\begin{table*}
\caption{Summary of previous studies on GRB precursors. The different requirements for identification of precursor emission imposed by each author and the conclusions reached in each paper are presented.}
\centering
\begin{tabular}{|c| c| l|}
\hline\hline
Paper & Requirements for Precursors & Conclusions\\
& (Definition) $^{*}$&\\
\hline
K95 & The quiescent interval between & (1) The properties of the main event are  \\
 	   & precursor and main event is & independent of the existence of precursors. \\
 	   & at least as long as the duration  	& (2) No evidence for different environment\\
 	   &of the main event.	& or different emission mechanisms.\\
\hline
L05 & Precursor emission does not 	& (1) Precursors have softer spectra than \\
	   & trigger the detector.	&  the main events.\\
	   & The flux decays before the& (2) Most  precursors show non-thermal \\
	   & onset of the main event.	& emission.\\
\hline
B08 & The precursor flux falls below & (1) Precursors and prompt emissions  \\
	   & the background level before & have similar spectral properties and \\
	   & the rise of the main event.& energetics.\\
\hline	
B09 & 															& (1) Similar spectral evolution.\\
	   &	same as in B08                    			& (2) Precursors may be produced by the \\
	   &															& same fireball mechanism as the main event.\\
\hline
       & same as in B08                               & (1) Precursors of short GRB  most likely \\
   \citet{2010ApJ...723.1711T}    &(only short GRBs \& short &are produced before the binary merger.  \\
& with extended emission)& \\
\hline
\end{tabular}
\label{table:Definitions} 

$^{*}$All the previous searches defined the main emission episode as the event with the highest peak rate, similar to our definition.

\end{table*}



The paper is organized as follows: In Section \ref{section:data}, we describe the extraction of light curves for the different samples. We detail the algorithm developed to search for precursor emission in Section \ref{section:algorithm}. In Section \ref{section:results}, we present our findings and compare them with the previous studies. We summarize our results and conclusions in Section \ref{section:conclusion}.

\section{Data}
\label{section:data}

We analyzed GRB light curves from the three main GRB catalogs: (i) the Gamma-ray Burst Monitor (GBM) on board Fermi Gamma-ray Space Telescope \citep{2009ApJ...702..791M} (ii) the Burst Alert Telescope (BAT) on the Swift satellite \citep{2004ApJ...611.1005G} and (iii) the Burst and Transient Source Experiment (BATSE) on board Compton Gamma-Ray Observatory \citep{Fishman89}. We analyzed GRBs detected prior to Jan. $1^{st}$, 2014. The search was confined to long GRBs with nominal duration $T_{90}$ $(T_{90}>2\,s)$, where $T_{90}$ is defined as the time interval during which 90\% of the GRB fluence was detected with $5\%$ fluence detected both before and after the interval. Note that $T_{90}$ was the only property considered in identifying long GRBs. The main reason for this selection is the reduced accuracy of the search for variability shorter than the bin size of the available light curves.

\subsection{Fermi-GBM}

Fermi-GBM consists of 12 NaI detectors, sensitive to energies from 8\,keV to $\sim$1\,MeV, which cover the entire unocculted sky, along with 2 BGO detectors sensitive to higher energy photons ($\sim$200\,keV to $\sim$40\,MeV). The GBM burst catalog\footnote{\url{http://heasarc.gsfc.nasa.gov/W3Browse/fermi/fermigbrst.html}} consists of 1276 GRBs (07/2008-12/2013) and the data are publicly available\footnote{\url{http://heasarc.gsfc.nasa.gov/FTP/fermi/data/gbm/}}.

For each GRB, we select the subset of the detectors that observed the burst, using the selection employed in the burst catalog (\citealt{2014ApJS..211...13V}). These detectors have good viewing angles of the source ($\leq 60^ \circ $) and are not blocked by the satellite. For each burst, we use Continuous Time data (CTIME) including all (8) energy channels, obtaining the detected count rate in 256\,ms bins. We use the observed count rate in a time window up to $[-1000, 750]$\,s around the GRB trigger time. We use smaller time windows if data are not available for this full interval.

Signals that are statistically significant in only one detector (e.g., phosphorescence spikes) are vetoed. 
The light curves are also inspected for telemetry gaps (periods during which the data are unavailable due to some technical issue), and the part before/after the discontinuity is discarded.
To avoid edge artifacts, we excluded from the analysis 30\,s at the edge of the time series, if the available data do not cover the full time interval $[-1000, 750]$\,s.

Bursts without available catalog files, or those which were only observed by 1 detector (42 GRBs), are not included. 
GRBs with inadequate data (less than 100\,s before/after the trigger time) and GRBs with multiple triggers within the interval of interest\footnote{If a time series contains multiple GRB triggers, one of the GRBs will be falsely identified as pre/post-peak emission with our automated analysis.} (22 and 9 GRBs, respectively) are also automatically excluded. After visual inspection of all light curves, we further exclude 39 bursts with significant, rapid background rate fluctuations to avoid spurious detection. Therefore, the final sample includes 956 GRBs, out of the 1069 long GRBs in the catalog.

\subsection{Swift-BAT}

Swift-BAT is a sensitive gamma-ray detector with a wide field of view (1.4 steradians) designed to provide GRB triggers with accurate localization. During the considered observation period up to the end of 2013 (12/2004-12/2013), Swift had detected 833 GRBs (\citealt{2011ApJS..195....2S}).

The data are retrieved from the public archive\footnote{\url{http://swift.gsfc.nasa.gov/archive/}} and processed with the standard data analysis Software (HEASOFT 6.14) using the online calibration library\footnote{\url{http://heasarc.gsfc.nasa.gov/ftools/}} \citep{1995ASPC...77..367B}. For each burst, a 64\,ms background subtracted light curve in the 15-350\,keV energy range is extracted with the automated GRB script (\emph{batgrbproduct}). The 64\,ms bins are combined to construct light curves with uniform time resolution of 256\,ms.
The $T_{90}$ of the prompt emission is taken from the summary output of the automated pipeline. 

From the sample, we automatically exclude 71 GBRs for which the online data are not complete (batgrbproduct is unable to analyze these GRBs), 65 short GRBs (60 according to the $T_{90}$ from the automated process and 5 additional from the burst catalog), 100 GRBs with inadequate data (less than 100\,s at either side of the trigger), and 14 GRBs where the change of the background rate would affect the analysis. The resulting Swift-BAT catalog contains 583 bursts.

\subsection{BATSE}

BATSE consisted of 8 Large NaI Area Detectors (LADs) covering the energy range of $\sim25$\,keV to $\sim2$\,MeV, and was able to observe the entire unobstructed sky. Over its nine years of operation (04/1991-05/2000), it detected 2702 GRBs\footnote{The BATSE Current Gamma-Ray Burst Catalog : \url{http://gammaray.msfc.nasa.gov/batse/grb/catalog/current/}}.

For each burst, the daily LAD discriminator data (DISCLA) are retrieved from the archive\footnote{\url{ftp://legacy.gsfc.nasa.gov/compton/data/batse/daily}}. The DISCLA data consisted of counts collected by each detector in 4 energy channels and in 1.024\,s time bins. For the light curves, the set of optimal detectors is defined following a similar approach to that for Fermi-GBM. First, detectors with angles $\leq 60^ \circ $ between the detector's normal and the source are identified. If this set is empty or contains only one detector, the two detectors with the smallest viewing angles are chosen. The
rates from the optimal set of detectors and from all energy channels are combined. We remove signals that are statistically significant in only one detector. The $T_{90}$ of the burst is taken from the online catalog.


From the search, we automatically exclude 661 GRBs for which $T_{90}$ is not available in the catalog, and the 500 short GRBs. From the remaining GRBs, the extraction of the light curves is not feasible for 10 GRBs (the trigger time coincides with a telemetry gap or the relevant DISCLA file is corrupted) and 207 GRBs do not have at least 100\,s at both sides of the trigger. These are also excluded from the analysis. After visual inspection of the light curves, we further discard 41 bursts for which the background count rate undergoes significant and rapid variation, and 1 GRB for which the end of the nominal $T_{90}$ falls outside of the interval of interest and coincides with a telemetry gap. We exclude 8 GRBs for which the background count rate shows strong deviations from the expected a Poissonian distribution. Our final BATSE samples consists of the remaining 1275 GRBs.

\section{Analysis}
\label{section:algorithm}
\begin{figure*}
\begin{center}
\resizebox{0.93\textwidth}{!}{\includegraphics{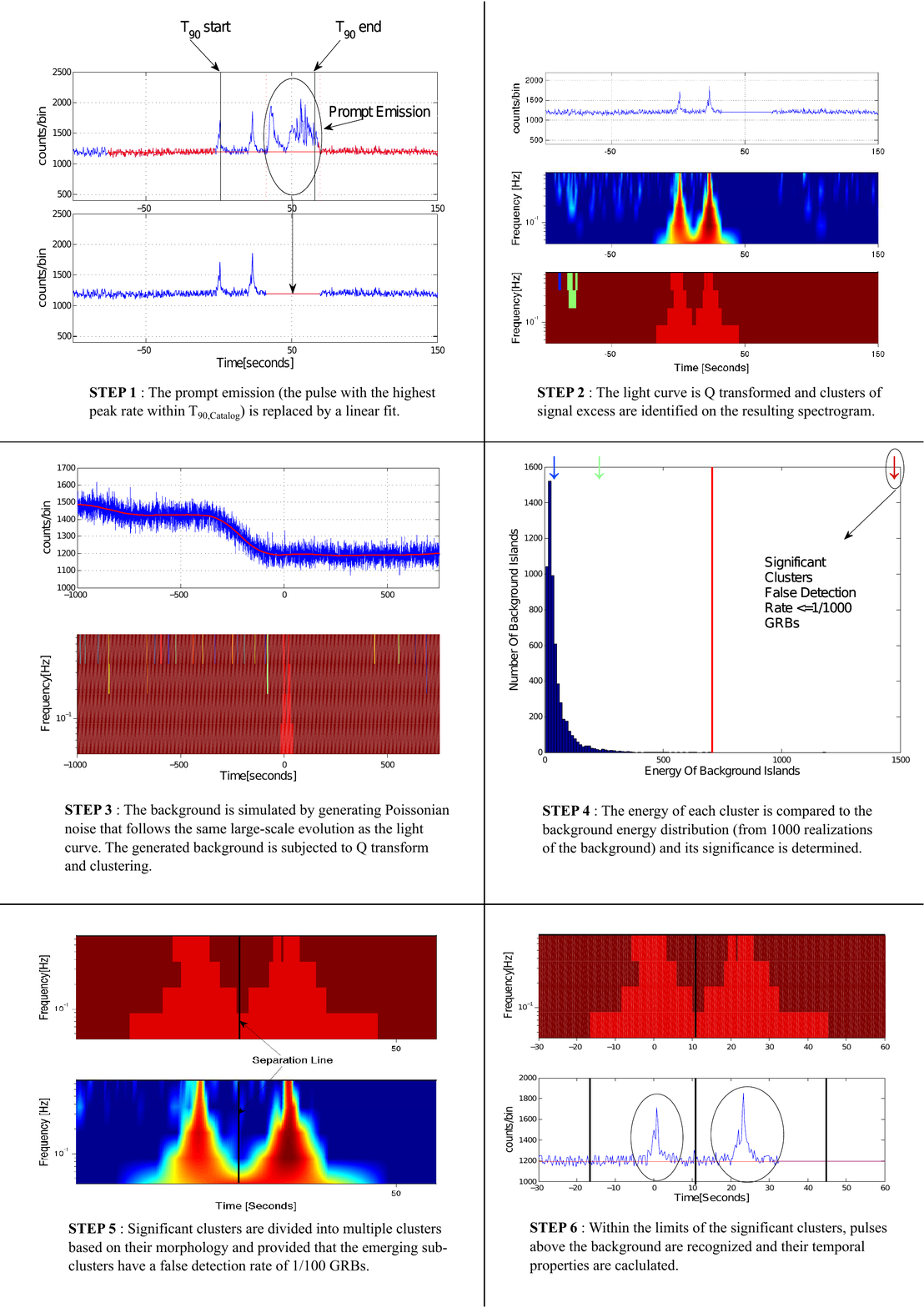}}
\end{center}
\caption{Steps followed in the automated algorithm developed for the search of isolated emission episodes.}
\label{Fig:Steps}
\end{figure*}

GRB time series are systematically analyzed with an automated algorithm to detect excess gamma-ray emission episodes. The steps of the automated data analysis are described below and illustrated in Fig.~\ref{Fig:Steps}.

\begin{figure}
\includegraphics[height=6.5cm,width=8.5cm]{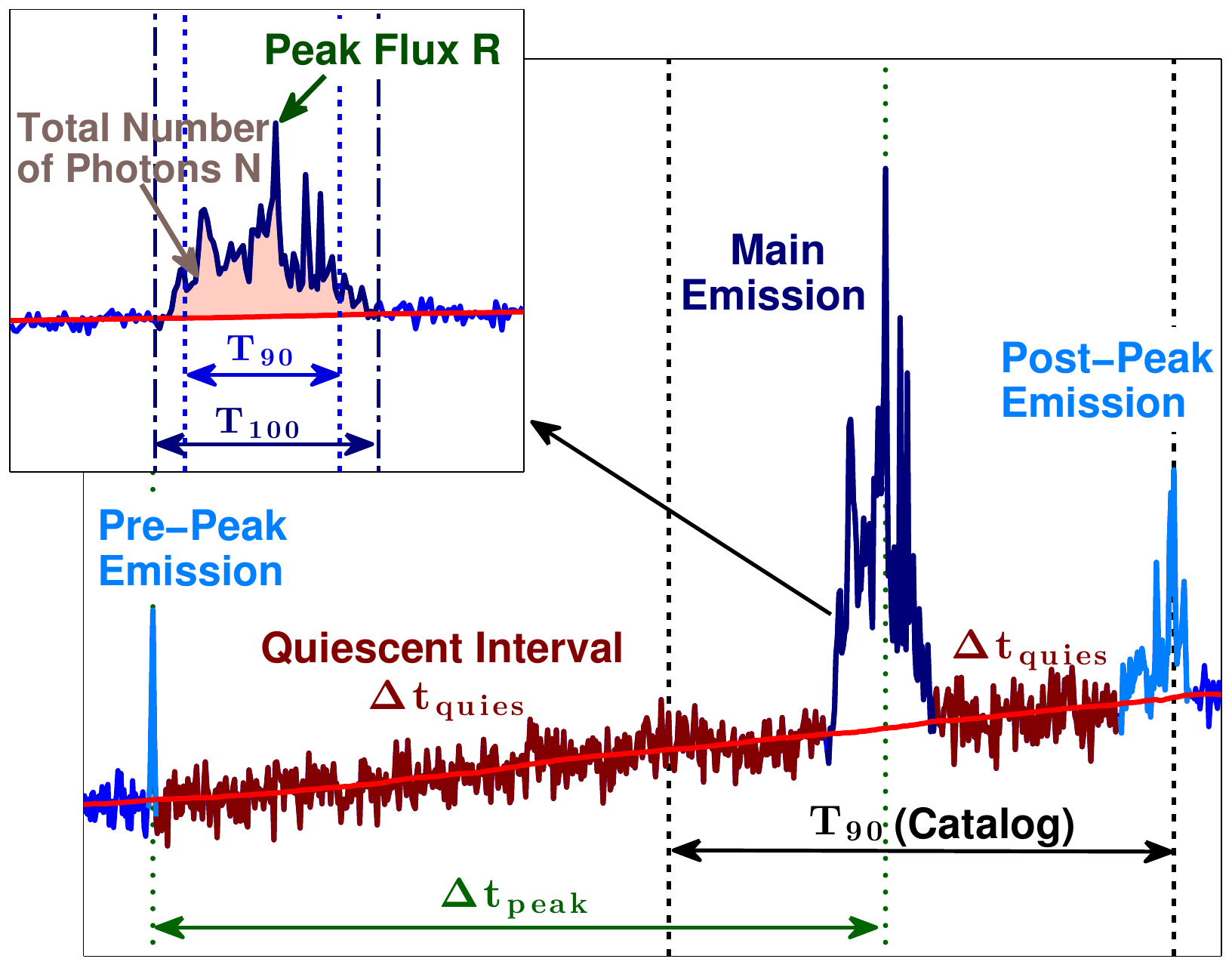}
\caption{Diagram illustrating the definition of isolated emission episodes and the temporal properties considered in the analysis (adopted from GRB 961125 detected by BATSE).}
\label{Fig:Definition}
\end{figure}

Of the emission episodes, the one with the highest peak gamma-ray flux, within the burst's nominal $T_{90}$ from the catalog, is identified as the main emission episode. We will call the emission periods prior to the main event pre-peak events, while those after the main event will be called post-peak. The definition is illustrated in Fig. \ref{Fig:Definition}.

First, we select the \emph{main emission episode}. 
We identify the boundaries of this episode as the closest times before and after the peak at which the detected rate of gamma-rays falls below the background rate for at least two consecutive time bins. This corresponds to an interval of 0.512\,s in Fermi and Swift, and 2.048\,s in BATSE.

After identifying the main emission, we remove it by changing the count rate during the episode to the background rate. We estimate the background rate by a linear fit over intervals adjacent to the main episode. For Fermi-GBM, the linear fit is calculated using the intervals around the main event that are indicated in the Fermi catalog to be used for background fits. For BATSE, the $[T_{90}^{\rm start}-50\,\mbox{s},T_{90}^{\rm start}-10\,\mbox{s}]$ and $[T_{90}^{\rm end}+10\,\mbox{s},T_{90}^{\rm end}+50\,\mbox{s}]$ intervals are used for the fit\footnote{The order of the polynomial fit and the intervals used to fit the light curves from BATSE were chosen to achieve the simplest, but still successful, automatic represantation of the background variation around the burst time.}. The light curves from Swift are background subtracted and the main burst is substituted with zero background level. (STEP 1 in Fig.~\ref{Fig:Steps}).

Next, we search for significant emission episodes in the time-frequency domain. We use a minimum temporal bin size of 0.256\,s for Fermi-GBM and Swift-BAT, and 1.024\,s for BATSE. Our analyzed frequency range is $[0.043\,\mbox{Hz},\,0.758\,\mbox{Hz}]$ for Fermi-GBM, $[0.014\,\mbox{Hz},\,0.776\,\mbox{Hz}]$ for Swift-BAT, and $[0.040\,\mbox{Hz},\,0.206\,\mbox{Hz}]$ for BATSE. The low frequency thresholds were selected to include the typical lowest expected variation of GRB light curves, while excluding the potential longer term variation of the background. The high frequency thresholds were chosen due to the finite temporal bin size.

We obtain the energy in time-frequency bins using an algorithm developed for gravitational-wave data analysis (Q pipeline; \citealt{2004CQGra..21S1809C}). 
In this method, time-frequency bins are tiled using bi-square windows with overlapping Gaussian enveloped sinusoids. 
The signal is whitened using linear prediction, after which a high-pass filter is applied. 
To ensure comparability between bins, the energy density in each bin is normalized to account for the varied window size and the tile overlapping.

On the resulting spectrogram, we identify clusters of neighboring tiles with excess energy compared to the background energy level. We set a threshold energy level at the 95$^{\rm th}$ percentile of the normalized spectral energy distribution over which a tile is considered significant and can be part of a cluster (STEP 2). Clusters that do not extend to tiles with at least two distinct frequencies are rejected, since astrophysical signals are expected to be variable in more than one timescales. Making use of the positive excess count rate expected from GRBs, we reject clusters during which the average count rate falls below the rate expected from the background. Following the identification of clusters, we assign a test statistic to each cluster that is the sum of the normalized spectral densities for the individual tiles within the cluster.

To calculate the significance of a cluster, we obtain the background distribution of the cluster test statistic. For Fermi-GBM and BATSE, we simulate Poissonian noise with mean that corresponds to the local background fit. Here, we approximate the background noise level as the 200-second moving average over the data (after the subtraction of the main emission episode). This duration is taken to be much longer than typical GRB durations, but shorter than the typical variability of the background average. For Swift-BAT, we use Gaussian noise with zero mean and standard deviation that corresponds to the local standard deviation of the normalized light curve. The local standard deviation for Swift is calculated with a 100-s moving average.  The simulated noise is subjected to the same analysis as the real data (STEP 3). The background distribution of the cluster test statistic is obtained over 1000 simulated realizations for each GRB. For GRBs with only partial data available in the $[-1000,750]$\,s interval around the trigger, the number of background realizations is increased to obtain the same overall background duration. The significance of a cluster is then taken to be the fraction of background GRBs with greater test statistic than the cluster in question (STEP 4). Clusters with a false detection rate less than or equal to 1/1000 GRBs with data spanning an interval of total 1750\,s are considered significant. The choice of the specific false detection rate is dictated by the sample size of the three catalogs to minimize the false detection of non-GRB-related signals.

Significant clusters, after identified, are removed from the light curve, and the analysis is repeated again. This repeated search is performed because the presence of significant clusters biases the normalization of the energy content of time-frequency bins. For the removal, the light curve during the time interval of the significant cluster is replaced with the expected background count rate. This removal and search process is repeated until no more significant clusters are found. 

Significant clusters are further separated if they include a short quiescent period. This can happen because the spectrogram at low-frequencies is affected by an interval of the light curve, and therefore two time intervals with excess energy can be connected in the time-frequency domain. We divide a cluster into two if 
(i) the two new sub-clusters contain tiles at higher frequencies than the highest-frequency tile at the quiescent period 
and (ii) both sub-clusters reach a significance corresponding to a false detection rate of at least 1/100 GRBs (STEP 5).

For the identified significant clusters, we determine the corresponding emission episodes similarly to the identification of the main emission episode above (STEP 6). We find the peak count rate within the time interval covered by the cluster, and find the closest times before and after the peak at which the count rate falls below the background count rate for two consecutive bins (0.512\,s of quiescence for Fermi-GBM and Swift-BAT, and 2.048\,s for BATSE).  This time interval, limited by the quiescent intervals at both sides, will be considered the duration of the emission episode ($T_{100}$). The total number of photons $N$ accumulated within $T_{100}$, and the peak flux $R$ of the episode are obtained. The $T_{90}$ of pre/post-peak emission is defined as the time interval within which 90\% (from 5\% to 95\%) of the photons are collected. The considered temporal properties of the emission episodes are presented in Fig. \ref{Fig:Definition}.

We note that a large fraction ($\sim 40\%$) of the light curves from Swift-BAT show significant variations of the background rate as a result of the re-pointing of the telescope. This affects the above analysis, as the significant change of the background level will mitigate the significance of emission episodes occurring at low-background intervals. To account for the changing background, these Swift-BAT light curves are analyzed in parts. First, the slew time is identified from the ``\emph{raw}'' light curve (before the background subtraction and normalization), as the time at which the background rate changes steeply. Then, each part of the light curve is analyzed separately, from identifying the fixed energy threshold for clustering (from the relevant part of the light curve) to calculating the properties of the emission episodes. If an event is recognized as significant around the slew time (i.e. its limiting times extend beyond the slew time), it is considered significant only if it is significant compared to the background distribution on both parts of the light curve. In 14 light curves, the telescope re-pointed more than two times. We exclude these GRBs from our analysis.


\section{Results}
\label{section:results}

We searched for isolated emission episodes in long GRBs observed by Fermi-GBM, Swift-BAT or BATSE.  When available, we searched within the time interval [-1000,750]\,s around the trigger time. Some of the GRBs do not have full data coverage. To eliminate any bias introduced from the different length of the time series (e.g., it is unlikely to identify isolated emission episodes, if the available data span a very limited time interval around the main event), for Fermi-GBM and BATSE we calculated the results of our search by considering the sub-sample of GRBs that have data in the interval of $[-500,500]\,s$ around the peak time of the main event and take into account only the data within this time interval. The light curves from Swift-BAT typically span only $\sim$200\,s before the trigger time. For Swift-BAT, we used the $[-200,200]\,s$ interval around the peak of the main event. These time intervals were chosen to contain a significant fraction of the bursts. While we prepared the statistics below with these subsets of bursts, our catalog contains all GRBs and data from the entire analyzed time interval.

\subsection{Statistics of isolated emission episodes and comparison with previous studies}

From the Fermi-GBM catalog, we analyzed 956 long GRBs, 864 of which have data within $[-500,500]\,s$. We identified isolated episodes, beyond the main events, in 216 GRBs with 94 bursts showing emission episodes before the main event. This corresponds to $8.9\%-13.2\%$ of the sample for a 95\% confidence interval. The 146 bursts with post-peak emission correspond to $14.9\%-22.2\%$ at 95\% confidence interval. More specifically, 77, 13 and 4 main events were preceded by 1, 2 and 3 isolated episodes, respectively. Similarly, 105, 32 and 2 main events were followed by 1, 2 and 3 post-peak episodes, respectively. Finally, individual bursts with 4, 5 and 6 isolated emission episodes following the main event were also identified.

The sample of GRBs detected by Swift-BAT with data in the [-200,200]\,s interval contained 464 GRBs out of 583 in the initial catalog, with 154 showing isolated emission episodes. We identified pre-peak emission in 64 of the GRBs, which corresponds to $10.8\%-17.3\%$ of the sample with 95\% confidence interval, with the main event being preceded by 1, 2, 3 and 4 emission episodes in 51, 8, 2 and 3 GRBs, respectively. We detected post-peak episodes in 104 bursts ($16.5\%-24.9\%$). More specifically, 85, 14, 3 and 2 bursts showed 1, 2, 3 and 4 post-peak emission episodes, respectively.

\begin{figure}
\includegraphics[height=6.5cm,width=8.5cm]{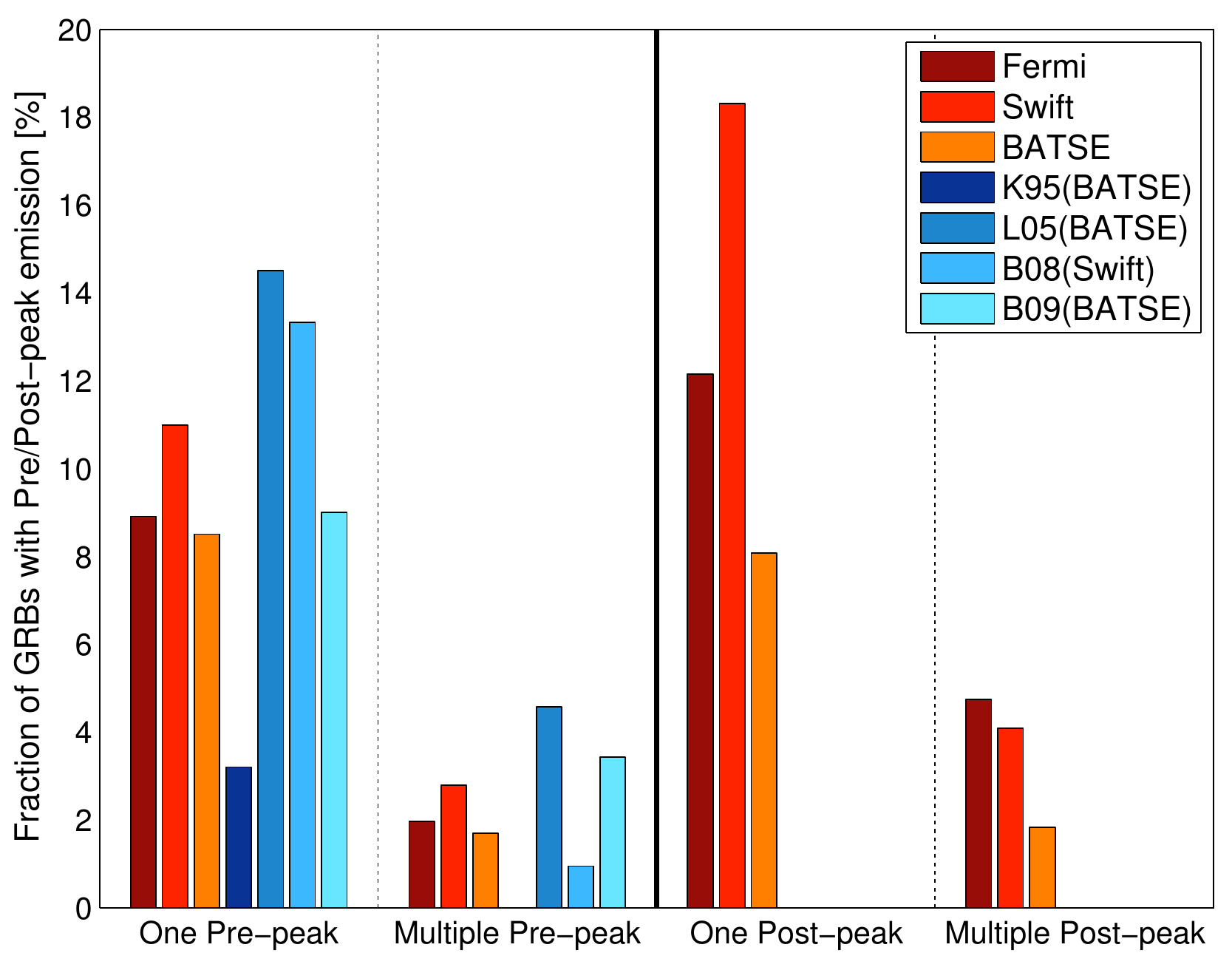}
\caption{Fraction of GRBs with pre-peak and post-peak emission episodes. The first three bars represent the occurrence of precursors in our sample (see legend). The subsequent bars illustrate the frequency of precursors in previous studies. On the right part of the graph, the frequency of post-peak events (which was not investigated previously) is shown.}
\label{Fig:PrecursorOccurence}
\end{figure}

\begin{figure}
\includegraphics[height=6.5cm,width=8.5cm]{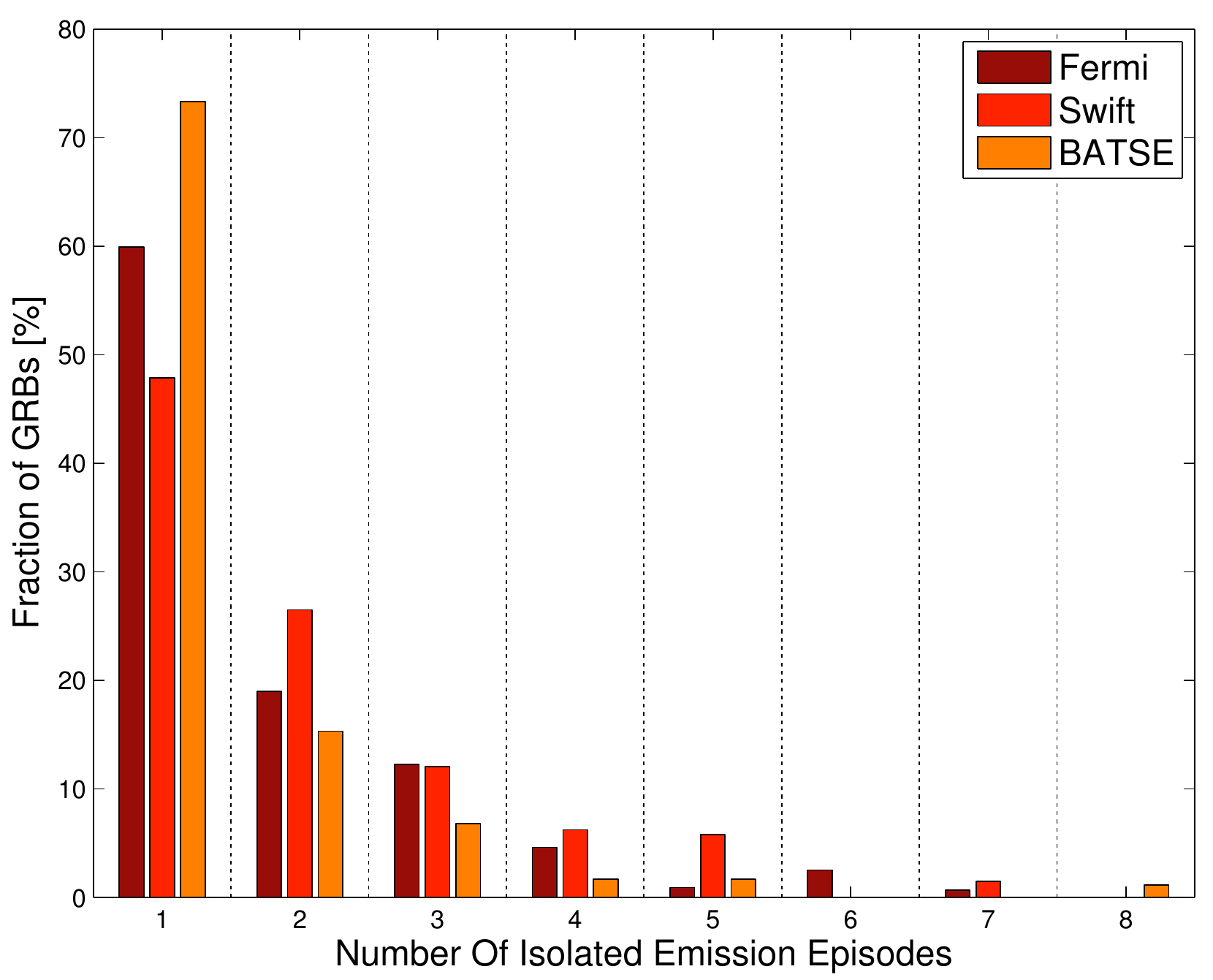}
\caption{Fraction of GRBs with isolated emission episodes versus the number of episodes in the three samples. 
The bar with one isolated episode corresponds to GRBs without pre-peak or post-peak emission.}
\label{Fig:NumberOfPulses}
\end{figure}

Of the 1275 GRBs in our BATSE catalog, 705 had full data coverage in the $[-500,500]\,s$ interval. Out of these, isolated emission episodes were detected in 134 bursts. In detail, we detected pre-peak episodes in 72 GRBs, which corresponds to $8.1\%-12.8\%$ of the sample with 95\% confidence. Out of these, 60, 8, 2 and 2 main-events were preceded by 1, 2, 3 and 4 emission episodes, respectively. The occurrence of post-peak events is $7.9\%-12.4\%$ (70 of the 705 GRBs show emission episodes after the main event), with 57, 11 and 2 main events followed by post-peak episodes, respectively. The above results are summarized in Fig.~\ref{Fig:PrecursorOccurence}. In Fig.~\ref{Fig:NumberOfPulses}, we show the distribution of the total number of isolated emission episodes of GRBs in the three different samples.

The percentage of GRBs with pre/post-peak events in all three samples is consistent (slightly higher in Swift and slightly lower in the BATSE sample). We note that the time interval under consideration, in the light curves from Swift, was more limited. It is not surprising that this did not have a great impact on the results, since we expect most pre/post-peak events to be close to the main events. For example, in the sample from Fermi only $\sim$4\%  (5 of the total 115) of the pre-peak episodes are identified 200\,s before the main event. The lower occurrence of isolated events in BATSE may be due to the data's limited time resolution compared to those of the other two missions. On the other hand, the number of identified Swift-BAT events is higher than, e.g., the number of Fermi-GBM events possibly due to the wider frequency range for Swift-BAT, allowed by the accurate background subtraction.

To further explore the effect of the difference between Fermi-GBM and Swift-BAT, we compared the GRBs that were detected simultaneously by these two missions. (BATSE has no contemporaneous data with Fermi or Swift.) From the 104 GRBs that were simultaneously detected by both instruments, 59 were included in the final samples; 31 GRBs showed neither precursor nor post-cursor emission. In 5 GRBs the same emission episodes were identified. From the remaining 23 bursts, in which the identification of emission episodes was not identical for the two missions,  the main conclusion is that our detection algorithm performed better on GRB’s detected by Swift-BAT than those detected by Fermi-GBM. This is likely due to the wider frequency range used for Swift, the better background subtraction and Swift-BATs higher signal-to-noise ratio for these GRBs. We also note that the two detectors are sensitive in different energy bands, with Swift-BAT covering the low energy range of Fermi-GBM. 


The topic of isolated pre-peak emission has been previously investigated in the context of GRB precursors (weaker emission episodes before the main event separated by a quiescent interval). In the sample from Swift, we identified pre-peak emission in $10.8\%-17.3\%$ of the sample, which is consistent with the percentage of identified precursors in B08 ($8.5-22.8\%$), in which a sub-sample of Swift GRBs with known redshift was used. In the BATSE sample, pre-peak emission was identified in $8.1\%-12.8\%$ of the GRBs. This exceeds the identified percentage from K95 ($2.1-4.8\%$) in which a strict requirement for the minimum quiescence was imposed (the quiescent interval between the precursor and the main event was at least as long as the duration of the main event). L05 identified precursor activity in $12.9-27.1\%$ of the sample. We note that L05 did not require a minimum interval of quiescence. Finally, B09 found precursor activity in $11.1-13.9\%$ of the bursts, consistent with our percentage. 

A careful one-to-one comparison of the previously identified precursors with our pre-peak emission episodes, yielded the following : A number of events identified before as precursors were not consistent with our requirement of minimum quiescence and were considered as part of the main event. We concluded that the identification of isolated emission episodes (or precursors) is sensitive to the definition. Therefore, the fact that our results are comparable with B08, B09 is not surprising given that a similar definition of precursors was employed. Moreover, a smaller fraction of precursors did not pass our significance threshold and was not considered a significant pre-peak emission episode in our sample. One other effect that can account for some of the differences is the set of detectors used for the construction light curves. K95 and L05 selected the detectors that maximize the signal, which in some cases may have resulted in precursors with stronger signal increasing their detection potential. The set of detectors for our search was chosen with an automated algorithm to avoid any human selection effects. Finally, our detection algorithm seems to be less sensitive to slow-rising light curves.

\subsection{Temporal properties of isolated emission episodes and correlations with the properties of the main emission}

We calculated the average properties of pre-peak and post-peak events from the different samples. The median and average $T_{90}$ of pre-peak, post-peak and main emission in the three samples, as well as the relevant number from previous papers are shown in Table \ref{table:results}. 
Moreover, we defined the temporal separation of pre/post-peak emission from the main event in two ways; $\Delta t_{\rm peak}$ is the peak-to-peak separation of the pre/post-peak events from the main episode and $\Delta t_{\rm quies}$ is the time difference between the end of a pre-peak event to the beginning of the main burst (or the end of the main event to the beginning of a post-peak emission). The two definitions are shown in Fig.~\ref{Fig:Definition}. 
The median and average $\Delta t_{\rm peak}$ for pre-peak and post-peak events for the three samples and previous studies are also included in Table \ref{table:results}.

\begin{table*}
\caption{Temporal characteristics of precursors and post-cursors. Results from previous studies are shown for comparison.}
\centering
\begin{tabular}{|c| c| c| c|c|c|c|c|}
\hline\hline
Sample & \# of GRBs  & Average &Median & $5^{th} - 95^{th}$ &Average &Median & $5^{th} - 95^{th}$ \\
& & $T_{90}$ [s] &  $T_{90}$ [s] & $T_{90}$ [s] & $\Delta t_{\rm peak}$ [s] & $\Delta t_{\rm peak}$ [s] & $\Delta t_{\rm peak}$  [s] \\
\hline	
Fermi main& 864 (Long) 	& 21 & 15& $3-63$ & & \\
\hline	
Fermi pre-peak& 864 (Long) 	& 13 & 8& $3-38$ & 52 &31& $5-181$ \\
\hline
Fermi post-post & 864 (Long)  & 13 &10 & $3-34$ &	82&39&$8-338$\\	
\hline		
 Swift  main & 464 (Long) & 23 & 15 &$3-67$&& \\	
\hline
 Swift  pre-peak & 464 (Long) & 11 & 9&$2-30$& 52&40&$9-170$\\	
\hline
Swift post-peak & 464 (Long)  & 17 & 11&$3-49$&54&42&$9-144$\\	
\hline 
 BATSE main& 705 (Long) & 32 & 25&$7-54$& &\\
\hline
 BATSE pre-peak& 705 (Long) & 20 & 15&$4-56$&107&92&$20-363$ \\
\hline
BATSE post-peak & 705 (Long)  & 27 &21&$4-66$&	112&69&$18-422$\\	
\hline		
K95 (BATSE)$^{*}$ & 748 & 12 &&&114& \\
\hline
L05 (BATSE)$^{**}$	& 131 (Long \& Bright) & 11& && 63&\\
\hline	
B09 (BATSE)$^{***}$	& 2121 & 15&&&50&\\
\hline
\end{tabular}

$^{*}$ K95 calculated the duration of precursors as the difference between the end and the beginning of an event ($T_{100}$), $^{**}$	L05 calculated the duration of precursors as the FWHM of the best-fitting Gaussian profile and the temporal separation of the events as $\Delta t_{\rm quies}$, $^{***}$B09 calculated the temporal separation of the events as $\Delta t_{\rm quies}$.
\label{table:results}
\end{table*}

The maximum $\Delta t_{\rm peak}$ observed, when the entire time interval ([-1000, 750]\,s) was considered, extended up to several hundreds of seconds. For example, the maximum $\Delta t_{\rm peak}$ of pre-peak emission in Fermi was $\sim$930\,s and $\sim$730\,s for post-peak. The maximum $\Delta t_{\rm peak}$ in the BATSE catalog was $\sim$850\,s for pre-peak events and $\sim$710\,s for post-peak events and the respective maximum values in the Swift catalog were $\sim$730\,s and $\sim$650\,s. We note that the maximum identified quiescent interval is limited by the available length of the time series. Quiescent intervals of comparable duration (and longer) are identified in some of the so-called ultra long GRBs (e.g., \citealt{2013arXiv1310.4944B}). The origin of these events is not clear. They may represent the tail of the duration distribution of long GRBs, (\citealt{2013ApJ...778...54V}; \citealt{2014ApJ...787...66Z}), while they were also suggested to have distinct progenitors (\citealt{2013arXiv1310.4944B}; \citealt{2013ApJ...766...30G}; \citealt{2013ApJ...779...66S}). We
note that some of the ultra-long GRBs were included in our search; however, part of their emission fell outside of the analyzed interval\footnote{The CTIME data from Fermi-GBM typically span [-1000, 1000]\,s around the trigger time}.

\begin{figure}
\includegraphics[height=6.5cm,width=8.5cm]{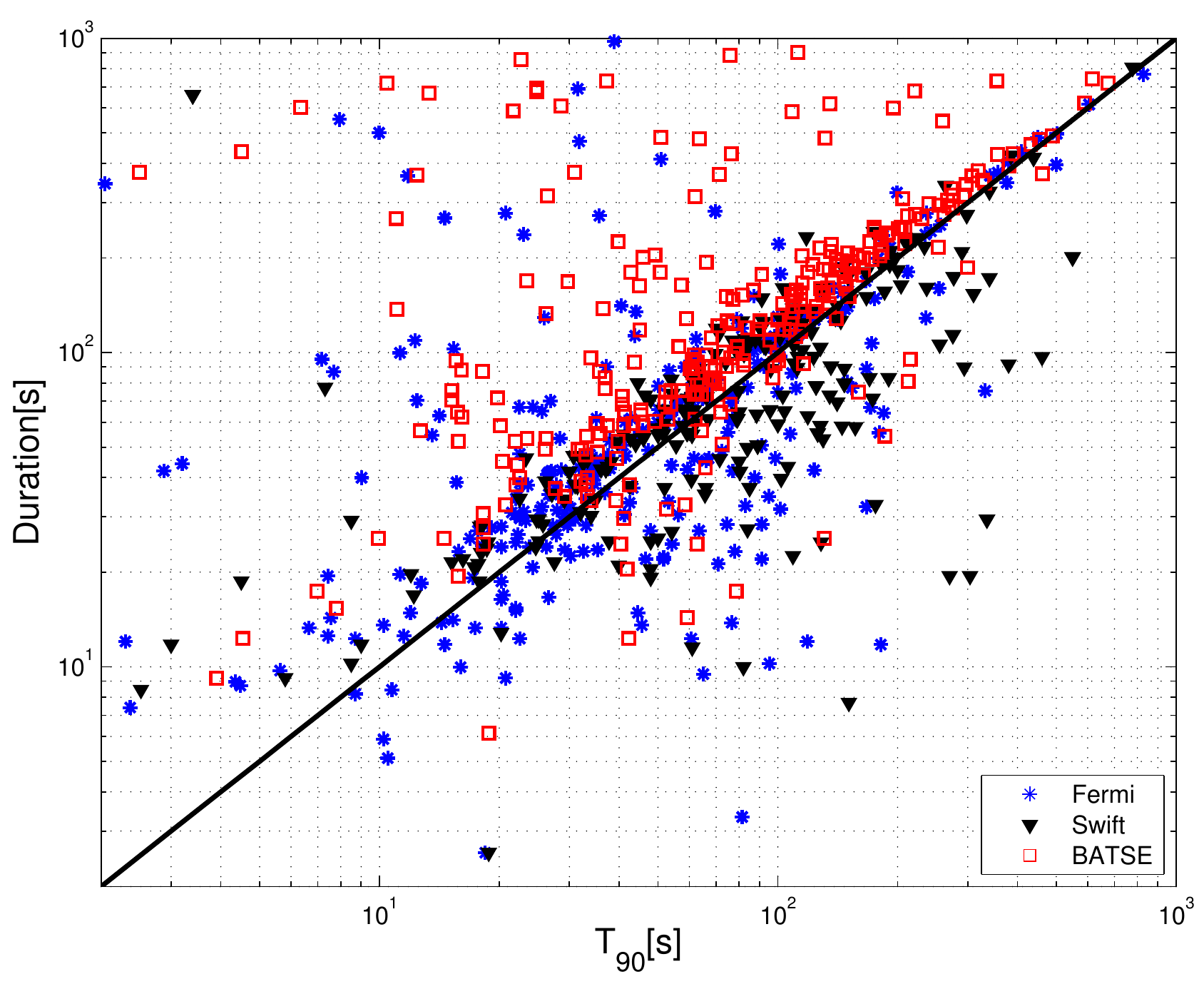}
\caption{Nominal $T_{90}$ of the burst versus the duration of the GRB, defined as the difference between the end of the last identified emission episode and the beginning of the first identified emission episode. Blue stars represent GRBs from Fermi, black triangles GRBs identified by Swift and red squares GRBs detected by BATSE. The black line represents the equality line.}
\label{Fig:T90VsDuration}
\end{figure}

In several cases, pre/post-peak emission episodes fall outside the nominal $T_{90}$ of the burst. This effect can be important for searches of non-electromagnetic signals from GRBs, in which the entire duration of the prompt emission needs to be taken into account\footnote{\citet{2014ApJ...787...66Z} argued that the central engine stays active for much longer than the burst's $T_{90}$ and suggested a new measure of the burst duration based both on gamma-ray and x-ray light curves.}. In Fig. \ref{Fig:T90VsDuration}, we compare the nominal $T_{90}$ of the burst to the total duration of the prompt emission of the burst, defined as the difference between the end of the last identified emission episode and the beginning of the first emission episode. A non-negligible fraction of points from all the samples lie above the equality line (i.e. the identified duration is much larger than the nominal $T_{90}$). The fact that most of BATSE points cluster around the equality line but slightly above must be viewed in light of the time resolution (bins of 1.024\,s) used for the calculation of duration in our search.

\begin{figure}
\includegraphics[height=7.5cm,width=8.5cm]{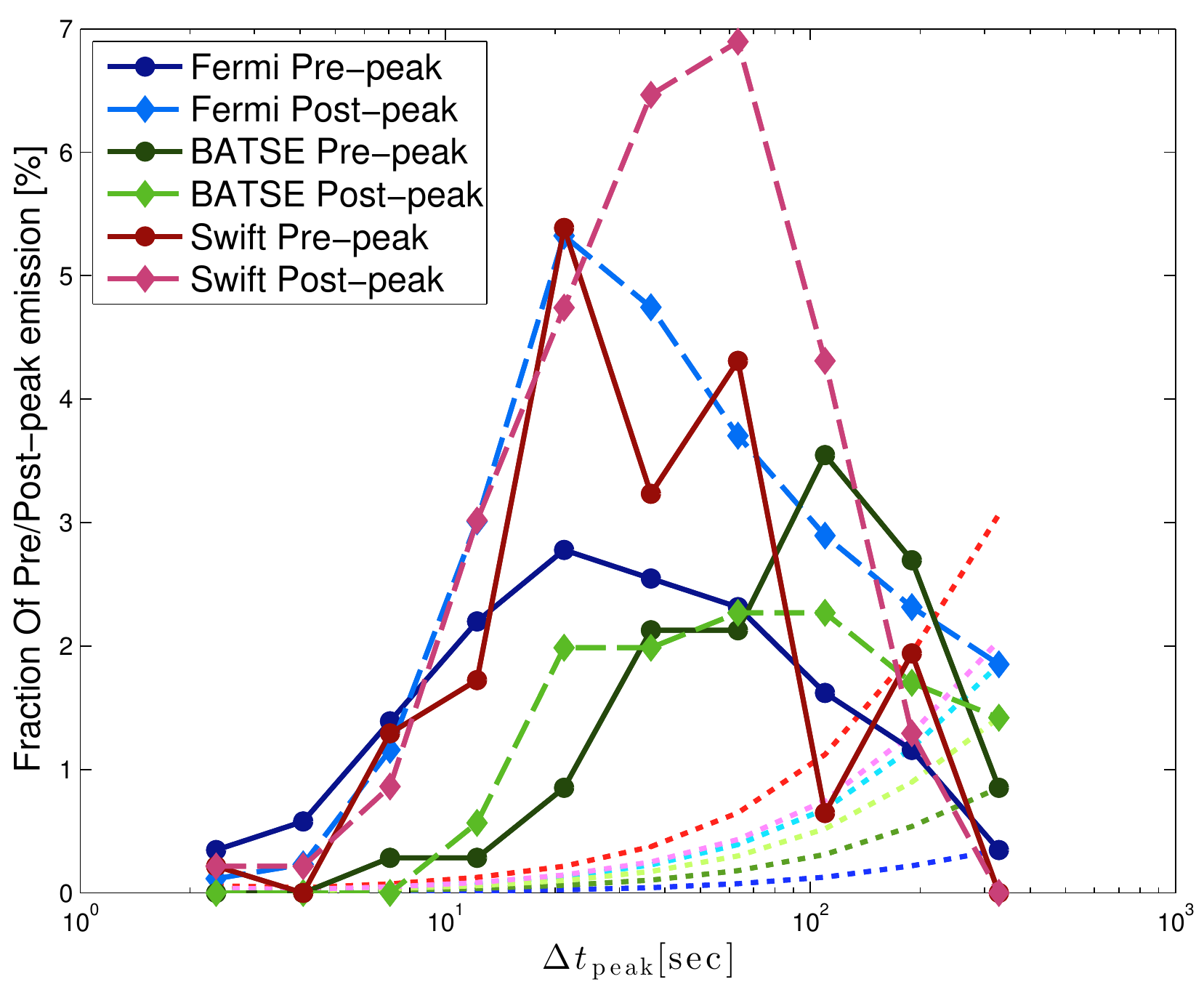}
\caption{Distribution of $\Delta t_{\rm peak}$ for pre/post-peak emission episodes. Dark blue, red, green dots represent pre-peak events in the Fermi-GBM, Swift-BAT and BATSE sample, respectively. Post-peak events in the relevant catalogs are illustrated with lighter colors and diamonds. The dotted lines represent the upper limit for the rate of non-astrophysical signals.}
\label{Fig:DelayHistogram}
\end{figure}

Furthermore, we present the distribution of $\Delta t_{\rm peak}$ for pre/post-peak events for the different samples in Fig.~\ref{Fig:DelayHistogram}. We note that the probability of falsely identifying a background feature as an emission episode is independent of the feature's peak time (i.e. the probability is the same regardless of the temporal separation from the main event) and thus the expected distribution of $\Delta t_{\rm peak}$ for false positives is uniform. We can therefore use the fraction of emission episodes at the farthest bin from the main event to set an upper limit on the false alarm rate of our search. This comparison shows that the large majority of the emission episodes identified by the search is likely astrophysical.

Next, we checked whether there was any correlation between the properties of the pre-peak emission episodes ($T_{90}$, $T_{100}$, number of photons $N$, peak flux $R$) and the respective properties of the main event using the Spearman's rank correlation coefficient. We did not find any significant correlation in any of the samples, consistent with the results of previous precursor studies. The properties of the post-peak events are also uncorrelated with the properties of the main event. One interesting finding is that pre/post-peak emission episodes carry a significant fraction of the total photon count of the bursts. 

\begin{figure*}
\begin{center}
\resizebox{0.9\textwidth}{!}{\includegraphics{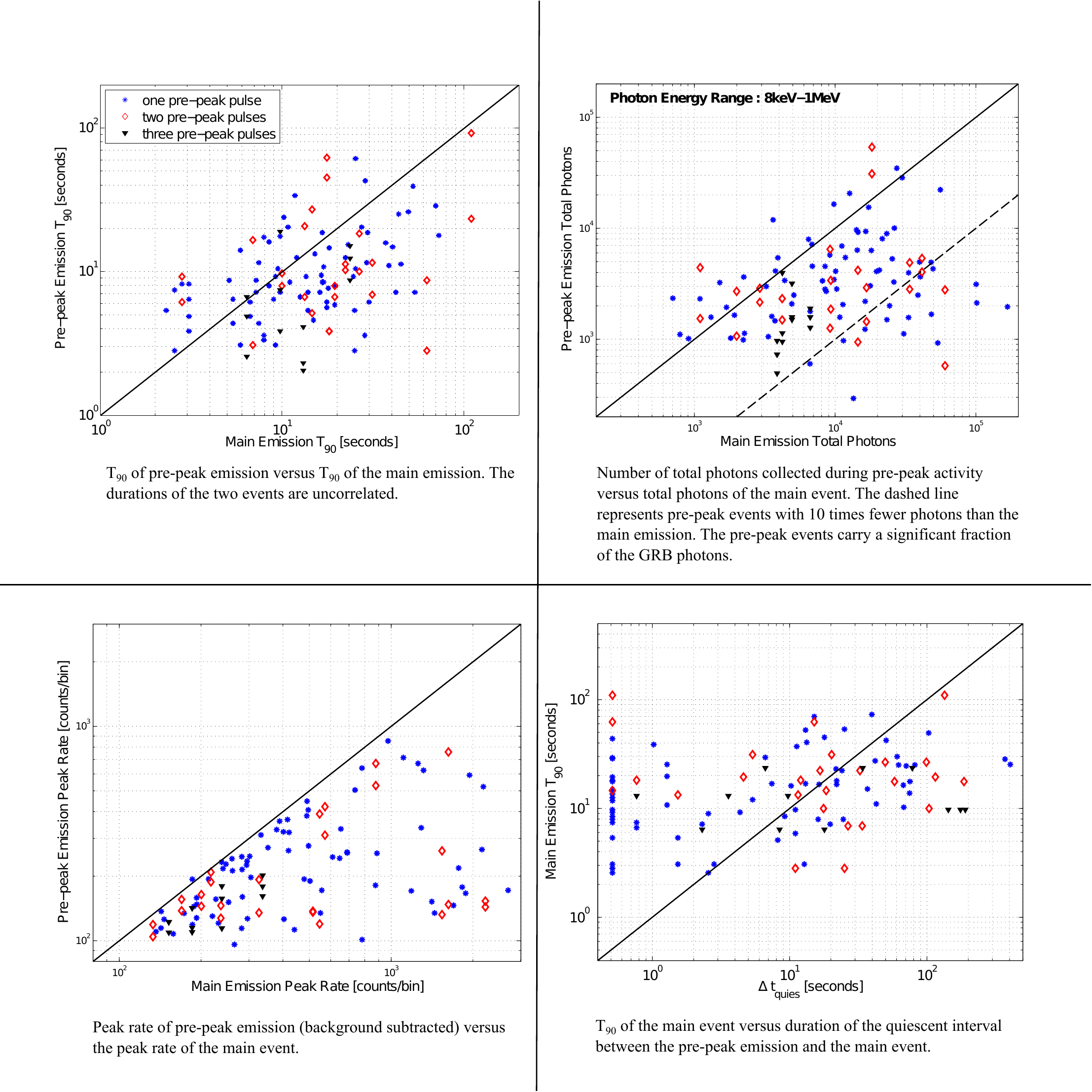}}
\end{center}\caption{Properties of pre-peak emission episodes versus the properties of the main event ($T_{90}$, total photons, peak rate) and correlation of the $T_{90}$ of the main emission with $\Delta t_{\rm quies}$. The color coding represents the number of pre-peak emission episodes observed. Blue stars for GRBs with one pre-peak event, red diamonds for GRBs with two pre-peak episodes and black triangles with three pre-peak events. The solid line is the equality line.}
\label{Fig:ResutlsSummary}
\end{figure*}

\cite{2001MNRAS.320L..25R} found that the duration of a quiescent period and the duration of the subsequent emission episode are correlated. To investigate this effect, we looked for correlations between the quiescent interval ($\Delta t_{\rm quies}$) and the duration of the events before and after quiescence. The duration of the main event as well as the duration of the identified isolated emission episodes are uncorrelated with the quiescent interval between the events. Finally, we checked if there is any correlation between the quiescent interval between successive episodes (without differentiating between pre-peak emission, main event and post-peak emission) and the duration of the previous or the following emission episode. The correlation reported in \cite{2001MNRAS.320L..25R} was not observed in our data. Our results for pre-peak emission identified in the sample of bursts detected by the GBM are summarized in Fig.~\ref{Fig:ResutlsSummary}.

\section{Conclusion}
\label{section:conclusion}

In this work, we presented a comprehensive catalog of GRBs with isolated emission episodes (pre-peak and post-peak emission). The catalog consists of a large sample of GRBs detected by Fermi-GBM, Swift and BATSE. We employed an automated identification method for an unbiased, large scale search. For the first time, a search for isolated emission episodes before the main event included GRBs detected by the GBM on the Fermi satellite. The search also included isolated emission episodes after the main event. For the first time, GRBs detected by different instruments were analyzed systematically with a common algorithm. Our findings from the analysis of three large datasets are summarized as follows:
\begin{itemize}
\item A significant percentage of GRBs ($>$10\%) show pre-peak activity. The percentage is similar in all three samples and is consistent with previous searches that employed a similar definition of precursors (B08/B09).
\item Post-peak activity was detected in a comparable, albeit slightly higher, fraction of GRBs $>$15\%.
\item The temporal properties (duration, peak flux, number of photons) of pre-peak and post-peak emission seem to be uncorrelated with the properties of the main event. They are scattered around the properties of the main burst, but no systematic trend was observed.
\item Pre-peak and post-peak events carry a significant fraction of the total photons of the GRB, regardless of the separation time of the two events.
\item Isolated emission episodes are separated from the main event by long quiescent intervals that extend at least up to $\sim$700\,s.
\item The properties of pre/post-peak episodes and the properties of the main event do no appear to be correlated to the duration of the quiescent interval.
\end{itemize}
\subsection{Future work}
Future additions to the presented analysis include the application of our automated search to identify isolated emission episodes in short GRBs. To resolve the temporal structure of short GRBs, light curves with time resolution shorter than the bursts' variability timescales are necessary, along with a frequency band extended to higher frequencies. Moreover, we will extend the catalog to incorporate GRBs detected by the interplanetary network (IPN; \citealt{2013ApJS..207...39H}).

It will also be interesting to use the analysis to study ultra-long GRBs. For instance, the distribution of the temporal properties of these GRBs (total duration or duration of quiescent intervals), may be informative on whether they have distinct progenitors from regular long GRBs (e.g., long separations between emission episodes have been interpreted as outliers in the distribution of quiescent intervals that possibly indicate a different underlying mechanism; \citealt{2002MNRAS.331...40N}; \citealt{2002A&A...385..377Q}).

The algorithm developed here is also suited for a robust statistical detection of untriggered GRBs and potentially ultra-long GRBs that were not identified. We note that we efficiently detected emission episodes that did not trigger the relevant detectors. Such searches were performed in BATSE (e.g.,\citealt{2001ApJ...563...80S}). Later, \citet{2005AstL...31..291T} detected pairs of spatially correlated bursts that were separated by long quiescent intervals and can be identified as ultralong GRBs.

\section*{Acknowledgement}
We are thankful to Neil Gehrels and Zsuzsa M\'arka for their useful comments that improved the manuscript. Also, we would like to thank Petr Kout for his help in making the catalog available online.


\end{document}